%% file: tess.tex
\documentclass[
    reprint,
    superscriptaddress,
    nofootinbib,
    amsmath,amssymb,
    aps,
    floatfix
]{revtex4-2}

\usepackage[T1]{fontenc}
\usepackage[utf8]{inputenc}
\usepackage{bm}          
\usepackage{upgreek}     

\usepackage{graphicx}
\usepackage{color}

\usepackage{chemist}
\usepackage{chemmacros}

\usepackage{comment}

\usepackage{lineno}
\usepackage{xcolor}

\begin{document}
\title{Controlled Spherulitic Crystal Growth from Salt Mixtures: A Universal Mechanism for Complex Crystal Self-Assembly} 

\author{Tess Heeremans}
    \email{Current affiliation: AMOLF, Science Park 104, 1098 XG, Amsterdam}
\affiliation{Institute of Physics, Van der Waals-Zeeman Institute, University of Amsterdam, Science Park 904, 1098XH, Amsterdam, The Netherlands}
\author{Simon Lépinay}
\affiliation{Institute of Physics, Van der Waals-Zeeman Institute, University of Amsterdam, Science Park 904, 1098XH, Amsterdam, The Netherlands}
\author{Romane Le Dizès Castell}
\affiliation{Institute of Physics, Van der Waals-Zeeman Institute, University of Amsterdam, Science Park 904, 1098XH, Amsterdam, The Netherlands}
\author{Isa Yusuf}
\affiliation{Institute of Physics, Van der Waals-Zeeman Institute, University of Amsterdam, Science Park 904, 1098XH, Amsterdam, The Netherlands}
\author{Paul Kolpakov}
\affiliation{Institute of Physics, Van der Waals-Zeeman Institute, University of Amsterdam, Science Park 904, 1098XH, Amsterdam, The Netherlands}
\author{Daniel Bonn}
\affiliation{Institute of Physics, Van der Waals-Zeeman Institute, University of Amsterdam, Science Park 904, 1098XH, Amsterdam, The Netherlands}
\author{Michael Steiger}
\affiliation{Department of Chemistry, Institute of Inorganic and Applied Chemistry, University of Hamburg, Martin-Luther-King-PLatz 6, 20146, Hamburg, Germany}
\author{Noushine Shahidzadeh}
    \email{n.shahidzadeh@uva.nl}
\affiliation{Institute of Physics, Van der Waals-Zeeman Institute, University of Amsterdam, Science Park 904, 1098XH, Amsterdam, The Netherlands}


 \begin{abstract}
Spherulites are complex polycrystalline structures that form through the self-assembly of small aggregated nanocrystals starting from a central point and growing radially outward. Despite their wide prevalence and relevance to fields ranging from geology to medicine, the dynamics of spherulitic crystallization and the conditions required for such growth remain ill-understood. Here, we report on the conditions to induce controlled spherulitic growth of sodium sulfate from evaporating aqueous solutions of sulfate salt mixtures at room temperature. We reveal that introducing divalent metal ions in the solution cause spherulitic growth of sodium sulfate.  For the first time, we quantify the supersaturation at the onset of spherulitic growth from salt mixtures and determine the growth kinetics. Our results show that the nonclassical nucleation process induces the growth of sodium sulfate spherulites at high supersaturation in highly viscous solutions. The latter reaches approximately 111 Pa$\cdot$s, triggered by the divalent ions, at the onset of spherulite precipitation leading to a diffusion limited growth. We also show that spherulites, which are metastable structures formed under out-of-equilibrium conditions, can evolve into other shapes when supersaturation decreases as growth continues at different evaporation rates. These findings shed light on the conditions under which spherulites form and offer practical strategies for tuning their morphology. 
\end{abstract}

\maketitle

\section*{Introduction}

\begin{figure}[!ht]
    \centering
    \includegraphics[width=0.5\linewidth]{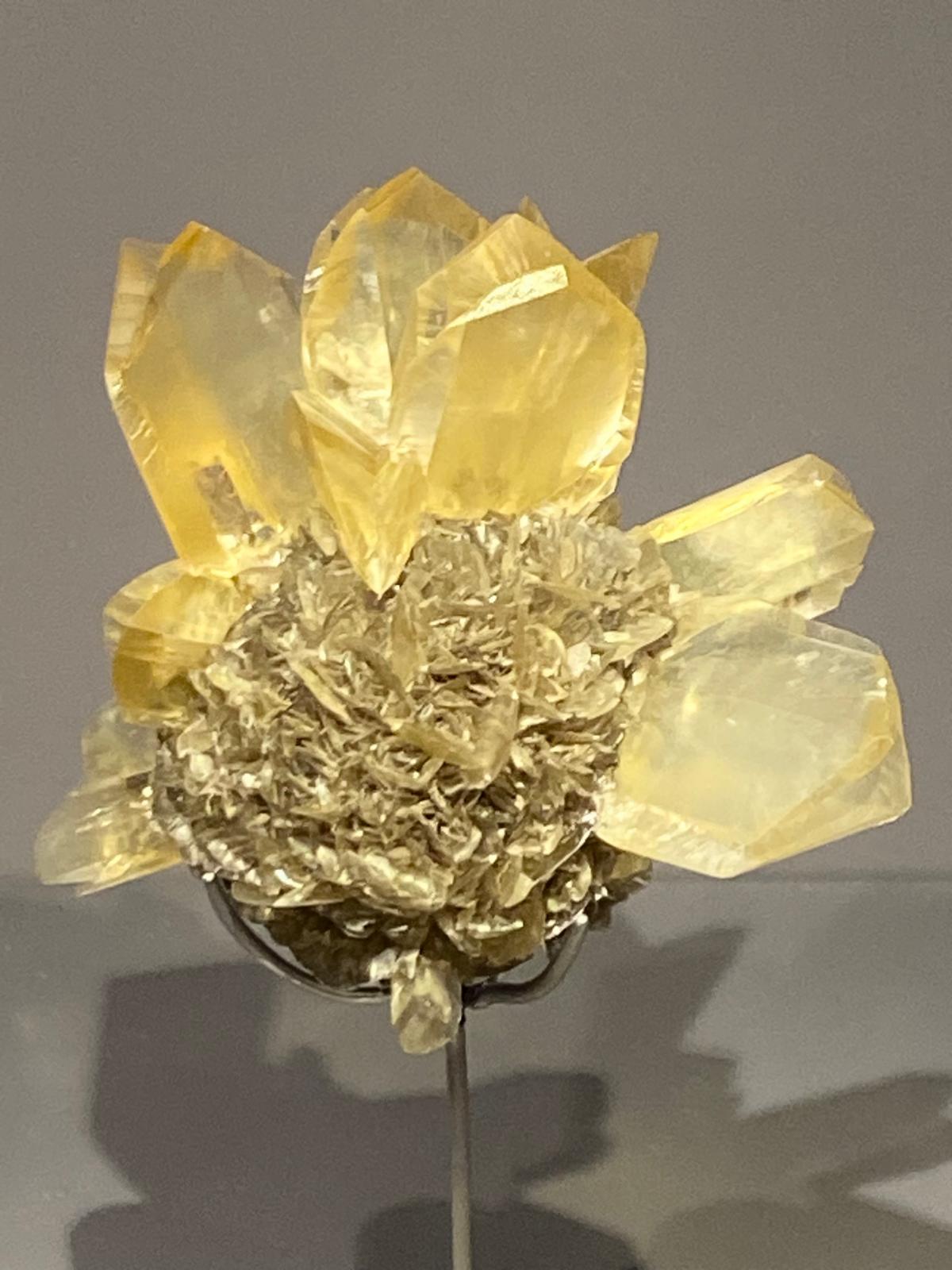}
    \caption{\textbf{Calcium sulfate spherulite found in Winnipeg, Manitoba, Canada}. Despite its complex appearance, it is composed solely of gypsum (calcium sulfate, CaSO$_{4}\cdot$2H$_{2}$O). Large amber-colored bladed  gypsum crystals can be seen on top of the polycrystalline gypsum macro spherulite. The bladed crystals grew during a secondary stage of crystallization. Collection of  the Gallery of Mineralogy and Geology in Paris, France.}
    \label{fig:CaSO4crystal}
\end{figure}

Crystallization involves the transition of ions, atoms or molecules from a disordered state (liquid or gas) to an ordered state (solid), where they arrange themselves into a regular pattern.  Although classical crystallization theories predict predictable morphologies based on the underlying crystal lattice and nature, complex chemical systems often produce intricate, unexpected structures that challenge our understanding \cite{garcia2017biomimetic}. Spherulites---radially symmetric polycrystalline aggregates formed by non-crystallographic branching \cite{granasy2005growth, Shtukenberg2012spherulites}---epitomize this complexity, emerging in diverse systems from geological formations to biological materials \cite{breitkreuz2021mineralogical, sokol2005cac2o4,al2010mechanism, krebs2004formation, wu2020naturally, sun2022growth}. Fig. \ref{fig:CaSO4crystal} shows a remarkable gypsum precipitate formed naturally from a complex mixture of minerals under unknown growth conditions. It consists of distinct morphological layers: small brown crystals near the precipitate's core, organized into a ball (macro-spherulite), and larger bladed crystals that likely grew on the surface of this ball.  
Mineralogists study meter-sized spherulites in rhyolitic lava \cite{breitkreuz2021mineralogical}, while in medicine, microscopic spherulites are associated with kidney stone formation and amyloid disorders such as Parkinson’s and Alzheimer’s diseases \cite{sokol2005cac2o4,al2010mechanism, krebs2004formation}. 
Despite this prevalence, existing theories do not adequately capture the dynamic interplay between chemical composition, environmental conditions, and crystal growth kinetics that drives these well-organized out-of-equilibrium morphologies \cite{beck_spherulitic_2010,wang2013confinement}. 

\begin{figure*}[!ht]
    \centering    
    \includegraphics[width=\textwidth]{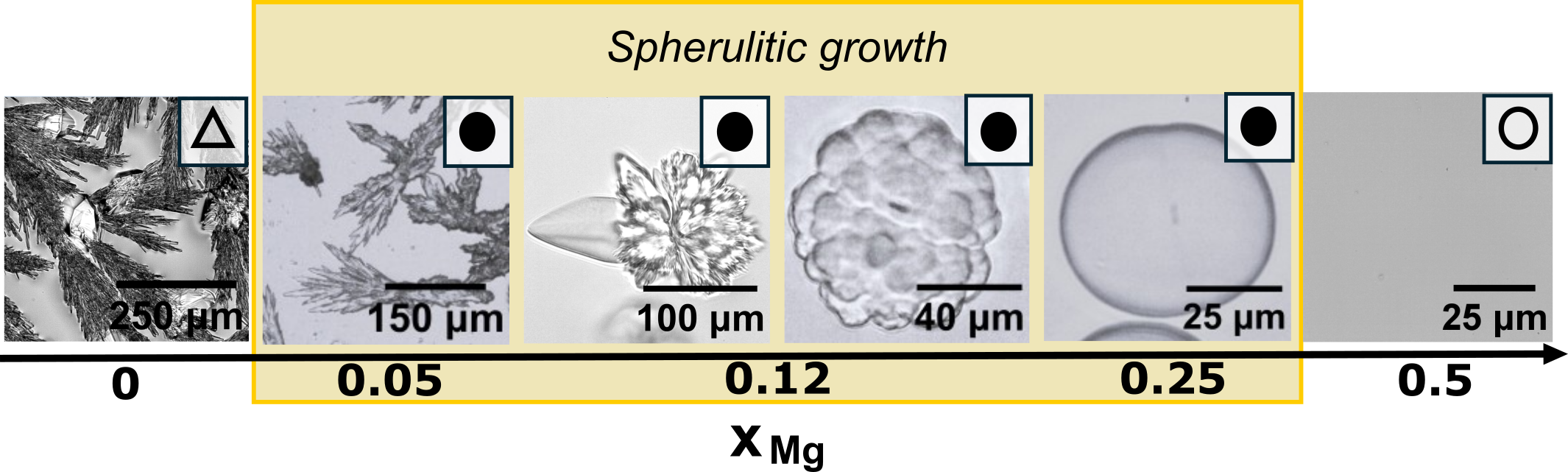}
    \caption{ \textbf{Microscopic images of precipitated sodium sulfate crystals  at different molar fractions of Mg.} Evaporation of unsaturated mixed salt solutions droplets with varying Mg molar fractions $x_\textrm{Mg}$ at $T=21\pm 1$\textdegree C resulting in three types of precipitation: i) polyhedral sodium sulfate crystals (labeled with triangles), ii) premature to full spherulitic morphologies (labeled with black dots), iii) evaporating droplets transforming into a gel-like state (amorphous state) (labeled with open circles).   \label{fig:x_ratio_OM}}
\end{figure*} 

Here we reveal that by mixing divalent sulfate salts with a monovalent (sodium) sulfate salt solution, controlled spherulitic growth of sodium sulfate with different structural behaviors emerges from evaporative mixed salt solutions. The universality of the process is verified in various mixed sulfate salt solutions. We also identify the range of mixing ratios leading to precipitation of spherulitic structures from mixed salt solutions. Our results from evaporating droplets of salt solutions show that sodium sulfate spherulites emerge from a two-step nucleation process at high supersaturation, followed by diffusion-limited growth kinetics as a result of the formation of a highly viscous salt solution upon evaporation. The supersaturation and local viscosity change at the onset of the spherulitic growth is determined. The impact of the evaporation rate on the evolution of the spherulitic structures after nucleation and on the final morphology of sodium sulfate polycrystalline self-assembly is also discussed.\\
Our findings not only provide valuable insights into the controlled growth of perfectly developed spherulites at room temperature but also open new avenues for industrial crystallization processes.

\begin{figure*}[!ht]
    \centering
    \includegraphics[width=\linewidth]{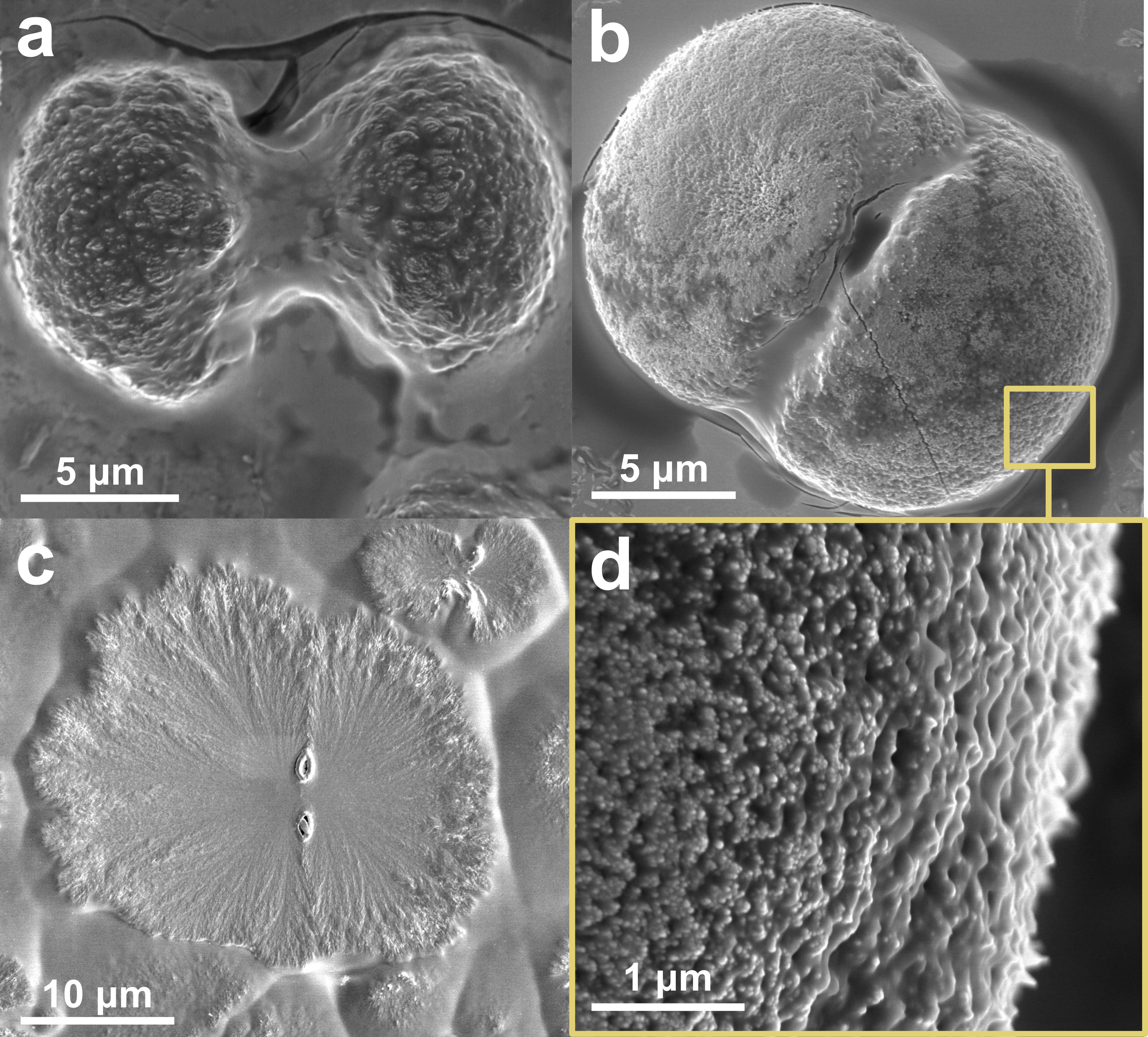}
    \caption{
    \textbf{Self-organized structure of sodium sulfate spherulites.} 
   Scanning Electron Microscopy images of spherulites grown in an evaporating droplet with iron molar fraction $x_\textrm{Fe}= 0.12$.  
    \textbf{a} premature spherulite at early stage of growth.
    \textbf{b} fully developed spherulite, 
    \textbf{c} Internal structure  of fully developed spherulite that grew at the edge of the droplet.
    \textbf{d} Zoomed-in view of the outer surface of the spherulite of panel b.}
    \label{fig:SEM}
\end{figure*}

\begin{figure*}
    \centering    
    \includegraphics[width=\textwidth]{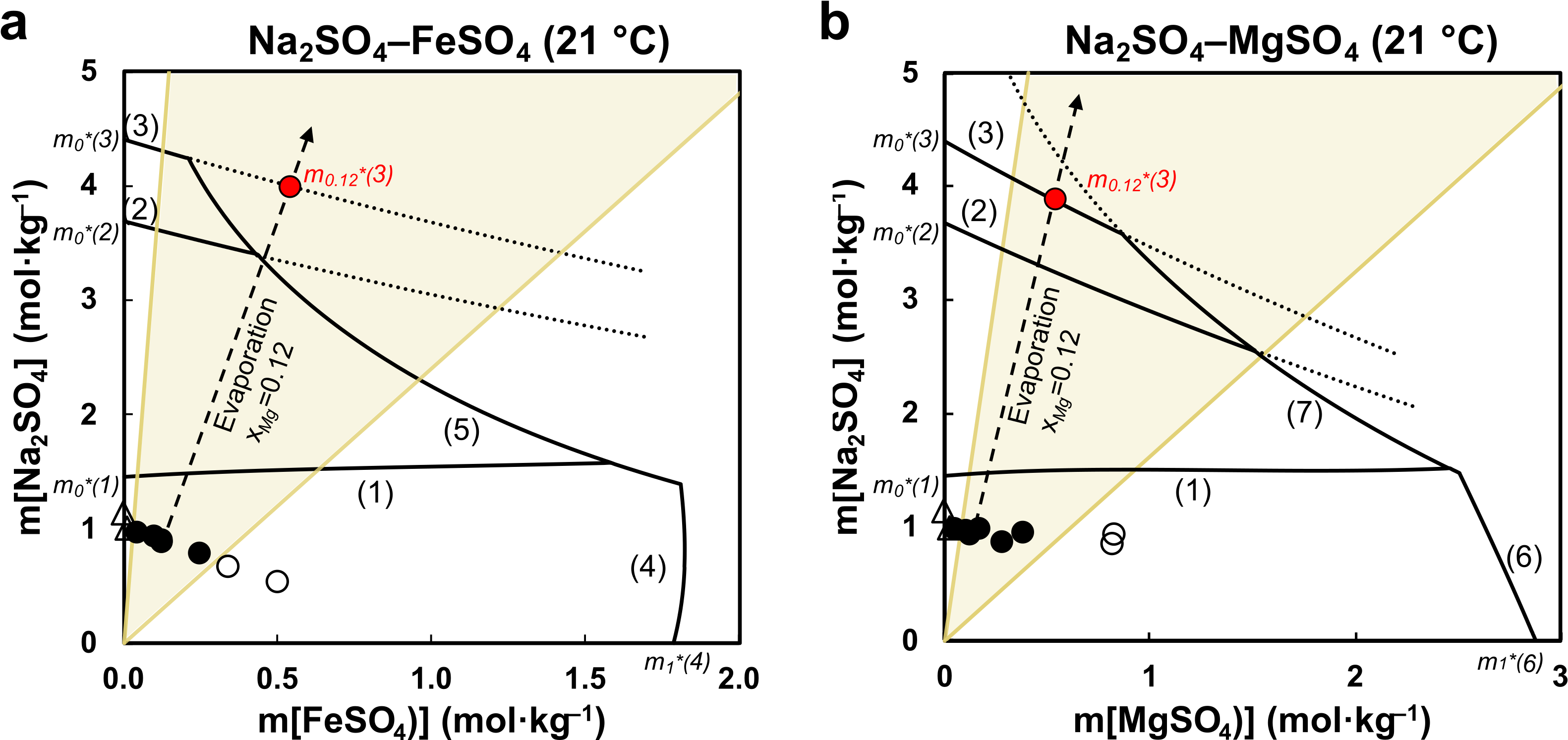}
    \caption{%
        \textbf{Morphodroms of spherulitic structures from mixed salt solutions.} Solubility isotherms at $T = 21\,^{\circ}\mathrm{C}$ in ternary systems: 
        \textbf{a} Na$_{2}$SO$_{4}$--FeSO$_{4}$--H$_{2}$O, 
        \textbf{b} Na$_{2}$SO$_{4}$--MgSO$_{4}$--H$_{2}$O.
       Black lines represent calculated solubilities by the molality-based model, accompanied by initial molal concentrations from evaporative droplet experiments shown as dots, triangles, and circles that correspond to Figure \ref{fig:x_ratio_OM}. 
        The phases are labeled as follows: 
        (1) Na$_{2}$SO$_{4} \cdot 10$H$_{2}$O (mirabilite); 
        (2) Na$_{2}$SO$_{4}$ (thenardite,  phase V); 
        (3) Na$_{2}$SO$_{4}$ (phase III); 
        (4) FeSO$_{4} \cdot 7$H$_{2}$O (melanterite); 
        (5) Na$_{2}$Fe(SO$_{4}$)$_{2} \cdot 4$H$_{2}$O (double salt bloedite-Fe); 
        (6) MgSO$_{4} \cdot 7$H$_{2}$O (epsomite); 
        (7) Na$_{2}$Mg(SO$_{4}$)$_{2} \cdot 4$H$_{2}$O (double salt bloedite-Mg). 
        The intersections at the y-axis (labeled as $m_{0}^{*}$(1), $m_{0}^{*}$(2), and $m_{0}^{*}$(3)) indicate the solubilities of the sodium sulfate phases without the introduction of divalent ions. 
        The ranges of \textit{x}-values where sodium sulfate spherulites precipitation is detected are highlighted in yellow: 
        for Na$_{2}$SO$_{4}$--FeSO$_{4}$--H$_{2}$O, $0.03 < x_\mathrm{Fe} < 0.29$, and for Na$_{2}$SO$_{4}$--MgSO$_{4}$--H$_{2}$O, $0.07 < x_\mathrm{Mg} < 0.38$. The dashed arrow indicates the trajectory of a drying mixed salt solution (x=0.12) upon evaporation crossing the solubility of phase III (red dot).
    } 
    \label{fig:steiger}
\end{figure*}

\section*{Results and discussion}

\subsection*{Divalent ion-induced self-assembly of sodium sulfate at the nanoscale}

Mixed sulfate salt solutions were prepared in Millipore water (\( \rho \approx 18.2\,\mathrm{M\Omega\,cm} \)) by mixing sodium sulfate Na$_{2}$SO$_{4}$ (Sigma Aldrich, purity > 99 \%), with iron sulfate heptahydrate FeSO$_{4}$$\cdot$7H$_{2}$O (Sigma Aldrich) or magnesium sulfate MgSO$_{4}\cdot$7H$_{2}$O (Sigma Aldrich) salts. For simplicity, the resulting solutions are referred to as Fe and Mg solutions. \\
Sodium sulfate is known to have two stable crystal phases: an anhydrous phase (Na$_{2}$SO$_{4}$, thenardite, phase V) and a decahydrated phase (Na$_{2}$SO$_{4}\cdot$10 H$_{2}$O, mirabilite) \cite{Wijnhorst2023}. In addition to these mineral forms a metastable dehydrated dendritic/needle-like phase III (Na$_{2}$SO$_{4}$) is known. Some laboratory experiments have reported another metastable phase that is hydrated by 7 water molecules (Na$_{2}$SO$_{4}\cdot$7H$_{2}$O). The solubility phase diagram of sodium sulfate in water, reported by Steiger \textit{et al.}\ \cite{steiger2008crystallization}, is provided in Fig. S\ref{fig:Appendix5_phasediagrams_literature}a.

We define the chemical composition of the resulting sulfate salt solutions system by the molar fraction $x$ of the newly introduced divalent ion X (Fe$^{2+}$, Mg$^{2+}$) as follows:
\begin{equation}
    x_{X}=\frac{n_{XSO_{4}}}{n_{XSO_{4}}+n_{Na_{2}SO_{4}}},
    \label{eq:Y}
\end{equation}
where $n_{XSO_{4}}$ is the molar amount of the divalent ion that is dissolved in the aqueous solution.
$x$ ranges from 0 (pure Na$_{2}$SO$_{4}$ solution) to 1 (pure FeSO$_{4}$ or MgSO$_{4}$ solution). 
Droplets (volume $V_{0}$ = 1 $\mu$L)  of these mixed salt solutions with various $x$ values were left to dry on cleaned glass slides at $T = 21 \pm 1^\circ$C and RH = 45 $\pm$ 5\%.  
Crystallization in the evaporating droplets with different values of $x_{Mg}$  or $x_{Fe}$ can be divided in three main categories, as shown in Fig. \ref{fig:steiger}: (i) precipitation of regular polyhedral-like crystals (black triangles), (ii) precipitation of premature and/or full spherulites (black dots), and (iii) precipitation of a gel-like (amorphous) state (open circles). In Na$_{2}$SO$_{4}$-MgSO$_{4}$--H$_{2}$O mixtures, spherulites form when 0.04 < $x_{Mg}$ < 0.38. For Na$_{2}$SO$_{4}$--FeSO$_{4}$--H$_{2}$O, spherulites form when 0.03 < $x_{Fe}$ < 0.29.
Fig. \ref{fig:x_ratio_OM} shows morphological states resulting from varying $x$ ratios in Na$_{2}$SO$_{4}$--MgSO$_{4}$--H$_{2}$O mixtures.  Low molar fractions ($x_{Mg}$ = 0.05--0.12) produce premature large spherulites with a well-defined needle-shaped crystal assembly with bladed crystals growing later in time from the assembly; this behavior is surprisingly similar to that of large bladed crystals grown on top of the gypsum sulfate spherulite in Fig. \ref{fig:CaSO4crystal}.
Higher molar fractions ($x_{Mg}$ > 0.12) produce  fully developed spherulites that are smaller in size, with different shapes from rough (raspberry shape) to very smooth (dense spherical shapes). Above a certain molar fraction ($x_{Mg}$ > 0.45),  spherulitic growth is fully suppressed and a bulk amorphous (gel-like) phase is formed. 
Experiments done with the Na$_{2}$SO$_{4}$--FeSO$_{4}$--H$_{2}$O mixed salt solution show similar morphological transitions in similar $x$ ratios. Scanning electron microscopy (SEM) images in Fig. \ref{fig:SEM} show the morphology of spherulites in a fully dried droplet of Fe solution ($x_{Fe}$ = 0.12); Fig. \ref{fig:SEM}a,b show a precipitated premature and fully developed spherulite at the center of the evaporating droplet. These  spherulites are covered by an amorphous (gel-like) state \cite{vancampenhout2025waterrichamorphousstatedrying}; in Fig. \ref{fig:SEM}c, the internal structure of the precipitated spherulites at the edge of the droplet can be seen: edge-grown spherulite are flattened because of the geometrical constraints imposed by the contact line of the evaporating droplet. Fig. \ref{fig:SEM}d zooms in on the surface of the spherulite. 
Chemical characterization by Raman confocal microspectroscopy reveals that the precipitated spherulites consist of an assembly of phase III nanocrystals. The overlapping Raman peaks (458, 616, 638, 996, 1076-1079, 1131, and 1197-1200 cm$^{-1}$) between the average spectra of 5 spherulites and the reference spectrum from phase III confirm this composition (Fig. S\ref{fig:raman_spherulites}).  To our knowledge, the ability of sodium sulfate to grow as spherulites has not been reported in the literature.  

Exploratory experiments involving the substitution of Fe$^{2+}$ and Mg$^{2+}$ with other divalent salts such as Cu$^{2+}$ and Zn$^{2+}$ also led to similar sodium sulfate spherulite precipitation (Fig. S\ref{fig:cu_spherulites}), suggesting that the mechanism of spherulitic growth is not system-specific and can occur in different mixtures of divalent sulfate salts.  The importance of divalence in triggering spherulitic growth is confirmed by the fact that the mixing of two monovalent sulfate salts (Na$_{2}$SO$_{4}$ and K$_{2}$SO$_{4}$) does not lead to the precipitation of sodium sulfate spherulites.

To determine the supersaturation at the onset of sodium sulfate spherulite precipitation, and to quantify their growth kinetics, we first establish the thermodynamic solubility phase diagrams for the ternary systems Na$_{2}$SO$_{4}$--FeSO$_{4}$--H$_{2}$O and Na$_{2}$SO$_{4}$--MgSO$_{4}$--H$_{2}$O. These phase diagrams help to determine the Na$_{2}$SO$_{4}$ phase solubilities in aqueous mixed sulfate solutions, and the formation of possibly new double salts depending on the molar ratio of FeSO$_{4}$ or MgSO$_{4}$ to Na$_{2}$SO$_{4}$, as defined by Eq. (\ref{eq:Y}).
 Using the experimentally validated molality-based model proposed by Steiger \textit{et al.} \cite{steiger2008improved,lindstrom2016crystallization} the solubilities of phases in the two ternary salt systems of Fe and Mg solutions are constructed at $T$ = 21$^\circ$C (other model parameters are listed in the method section). The model uses Pitzer equations \cite{pitzer2018ion} to calculate activity and osmotic coefficients, determining thermodynamic solubility products of solid phases. These equations compute excess Gibbs energy in electrolyte solutions, from which activity and osmotic coefficients are derived as functions of composition and temperature. The resulting ternary phase diagrams are presented in Fig.~\ref{fig:steiger}a,b. At $x$ = 0,  initial solubilities $m_{x}^{*}$ of mirabilite (labeled $m_{0}^{*}(1)$ in the figure), thenardite ($m_{0}^{*}(2)$), and phase III ($m_{0}^{*}(3)$) at $T$ = 21$^\circ$C correspond to the intersections with the y-axis (see Supplementary Table \ref{Table_ThermoSalt}).
By introducing divalent ions (Fe$^{2+}$ or Mg$^{2+}$) these solubilities change with increasing $x$ in the Na$_{2}$SO$_{4}$--FeSO$_{4}$--H$_{2}$O and Na$_{2}$SO$_{4}$--MgSO$_{4}$--H$_{2}$O systems, respectively. Notably, the solubilities of thenardite phase V (labeled (2) in Fig.~\ref{fig:steiger}) and phase III (3) decrease in the presence of divalent ions, while the solubility of mirabilite (decahydrate form) (1) slightly increases. The solubility line for the formation of double salt bloedite--Fe (Na$_{2}$Fe(SO$_{4}$)$_{2}\cdot$4H$_{2}$O)  (labeled (5) in the figure) and bloedite--Mg (Na$_{2}$Mg(SO$_{4}$)$_{2}\cdot$4H$_{2}$O (7) becomes relevant in these salt mixtures. 
In the equilibrium phase diagram, it can also be noted that in pure solutions of FeSO$_{4}$--H$_{2}$O or MgSO$_{4}$--H$_{2}$O (\textit{i.e.}, $x$ = 1), melanterite (FeSO$_{4}\cdot$7H$_{2}$O) (4) and epsomite (MgSO$_{4}\cdot$7H$_{2}$O) (6) are stable at $T$ = 21$^\circ$C with initial solubilities $m_{1}^{*}(4)$  and $m_{1}^{*}(6)$  (indicated on the x-axis). 

Experimentally determined regions corresponding to sodium sulfate spherulite precipitation (shaded yellow in Fig. \ref{fig:steiger}) are added to the equilibrium phase diagrams of both ternary systems by: 
\begin{equation}
m_{Na_{2}SO_{4}} = \frac{1-x_X}{x_X} m_{XSO_{4}}
\label{eq:inverseY}
\end{equation} 
where $X$ denotes the divalent ion (Fe$^{2+}$ or Mg$^{2+}$).

For a given ratio,  thermodynamic equilibrium phase diagrams can predict which phase should precipitate with the evaporation of water and the increase of the ion concentration. The increase in ion concentration is characterized by measuring the supersaturation (see method section). We express the supersaturation $\beta$ as the ratio of the molality of total sulfate in the supersaturated solution ($m_{S}$) at the onset of  precipitation and the saturated concentration ($m_{S}^*$) for the precipitation of sodium sulfate phase III  (see Fig. \ref{fig:steiger}, point $m_{0.12}^*$), respectively, i.e. $\beta = m_{S}/m_{S}^*$.  For, $x = 0.12$, the concentrations at the onset of spherulitic nucleation reach far beyond the field of view in Fig. \ref{fig:steiger}. We find $\beta_{Mg} = 1.4 \pm 0.3$ for the Na$_{2}$SO$_{4}$--MgSO$_{4}$ mixture and $\beta_{Fe}>3.3$  for the Na$_{2}$SO$_{4}$-–FeSO$_{4}$ system.
At the onset of precipitation the concentration is above the saturation concentration of all three sodium sulfate phases (mirabilite, thenardite, and phase III) as well as the double salt formation. However, we observe the precipitation of sodium sulfate phase III  within the spherulites region (in yellow) confirmed with confocal Raman chemical characterization of nanocrystals in spherulites. 


\subsection*{Growth mechanism driven by viscosity changes}

\begin{figure*}
    \centering
    \includegraphics[width=\textwidth]{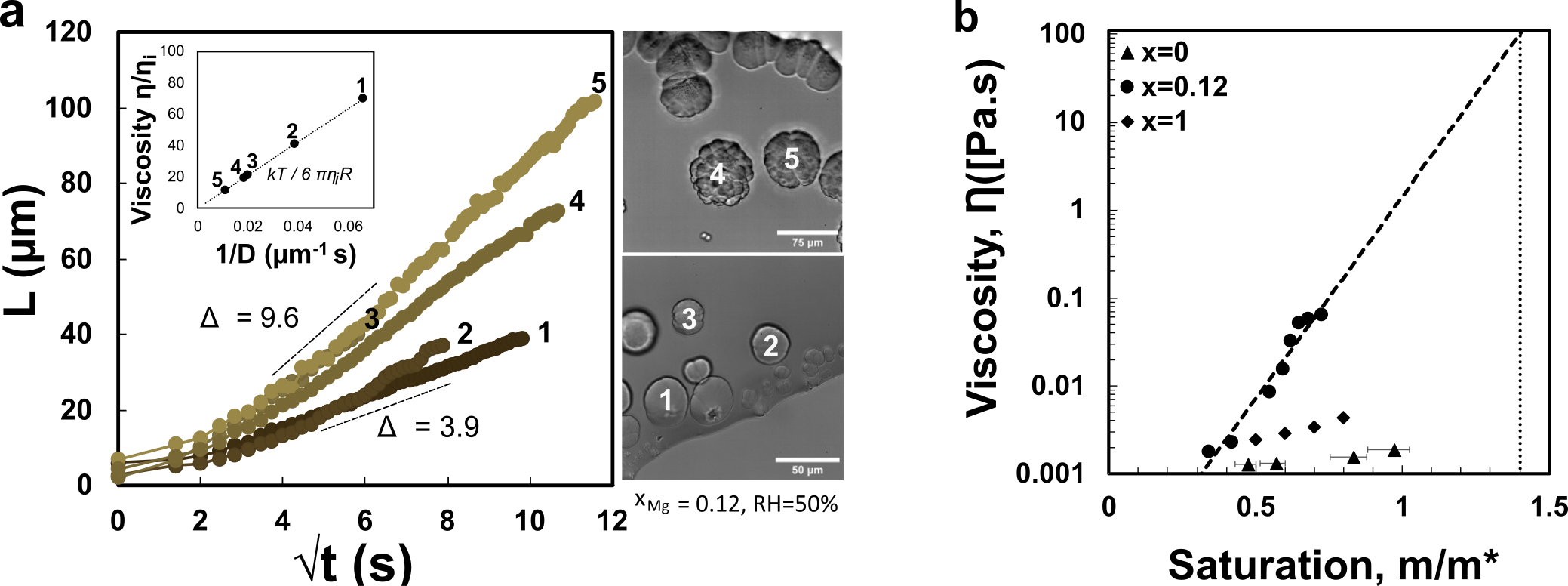}
    \caption{\textbf{Growth kinetics of spherulites in viscous mixed salt solution} \textbf{a} Spherulite size $L$ as a function of the square root of time $t$ at RH = 50\%. Each dataset is for one spherulite as shown in the photographs on the right: spherulites 1 and 2 are close to the contact line while 3, 4 and 5 are closer to the center of the droplet.  The linear trend indicates that the growth of the spherulites is controlled by diffusion, with spherulites further away from the edge of the droplet growing faster than those near the center (indicated by slope $\Delta$). The inset plot illustrates the linear relation between solvent viscosity $\eta$ and the inverse diffusivity $1/D$. A 12-fold viscosity increase is estimated for spherulite 5, and a significantly higher viscosity increase (70-fold) near the droplet's edge (spherulite 1).  \textbf{b} Rheological measurements of the saturation of sodium sulfate solutions with different molar Mg factions $x$. All measurements at $T$ = 21$\pm$1\textdegree C. While the viscosity $\eta$ of the pure sodium sulfate and pure magnesium sulfate solution are in the same order of magnitude as the viscosity of water, the viscosity of the $x_{Mg}=0.12$ solution as a function of the saturation follows an exponential trend. The dashed line indicates $\eta=4\times10^{-5} \textrm{exp}(10.6\beta_{Mg}$) with $\beta_{Mg} = m_{Mg}/m^*_{Mg}$ ($R^2 = 0.88$). Limited by the ability to dissolve, the viscosity is measured up to a saturation of $\beta_{Mg}$ = 0.73. By then, the viscosity is about 65 times higher than the viscosity of water. The exponential fit predicts that the viscosity is increased to 111 Pa$\cdot$s when spherulitic growth is initiated at $\beta_{Mg} = 1.4$ (dotted line).}
    \label{fig:growth_viscosity}
\end{figure*}

Typically, high supersaturations induce rough morphologies like dendrites or spherulitic growth \cite{sunagawa2007crystals} depending on the growth mechanism. To study the growth mechanism we follow the growth as a function of time of spherulites after precipitation, at different locations in the droplets  (Fig. \ref{fig:growth_viscosity}a). Interestingly, the spherulites grow in a diffusion-limited fashion, \textit{i.e.} proportional to the square root of time; the growth rates vary between spherulites located in the droplet periphery (lower growth rates) and spherulites closer to the droplet center (higher growth rates). Fe-containing solutions show growth characteristics that are remarkably similar to Mg-containing solutions, with comparable morphology and growth rates (Fig. S\ref{fig:Supp_Fe_growth}).
These results highlight spherulites as complex, dynamically evolving structures whose formation depends on intricate interactions between chemical composition, environmental conditions, and crystallization kinetics. Key factors influencing growth include hydrophilic surface effects and droplets contact angle ($\sim$ 30$^\circ$); the lower the contact angle, the more confined the solution at the droplet edges. Consequently, there will be a gradient in ion concentration and viscosity from the center to the periphery of the droplet \cite{shahidzadeh2015salt}.
We therefore investigate the evolution of the viscosity of the mixed salt solution as it evaporates. 

From the growth measurements in Fig. \ref{fig:growth_viscosity}a the increase in solvent viscosity can be estimated using the Stokes-Einstein relation, which states that the diffusion constant $D$ is inversely proportional to the solvent viscosity $\eta$, \textit{i.e.},  $D \sim 1/{\eta}$. The slope $\Delta$ of the spherulite growth measurements in Fig. \ref{fig:growth_viscosity}a, corresponds to $\sqrt{D}$ (in $\mu$m/s$^{1/2}$), from which we infer that 15.2 $\mu$m$^2$/s $\leq D \leq$ 92.2 $\mu$m$^2$/s. By the following relation, the solvent viscosity is estimated:
\begin{equation} \frac{\eta}{\eta_{i}}= \frac{D_{i}}{D}=\frac{kT}{6\pi\eta_{i}R} \frac{1}{D}, \end{equation} where $D_{i}$ is the initial diffusion constant of sulfate in water, reported as $1.07 \times 10^{-9}$ $\text{m}^2/\text{s}$ \cite{phreeqc3}. The inset plot in Fig. \ref{fig:growth_viscosity}a shows the linear relationship between the increased solvent viscosity and the inverse diffusivity. A 12-fold viscosity increase is estimated in the bulk region (spherulite 5), and a 70-fold viscosity increase near the droplet edge (spherulite 1).

 Furthermore, we perform rheological measurements on various magnesium sulfate solutions ($x_{Mg}$ = 0.12) with reduced water fractions up to a saturation of $\beta_{Mg}$ = 0.73. This approach simulates the stages of evaporation during which the concentration of ions increases progressively with the evaporation of water in the initial solution of the mixed salt solution (Fig. \ref{fig:growth_viscosity}b). The results show an exponential viscosity trend:
$\eta=4\times10^{-5} e^{10.6 \beta_{Mg}}$ ($R^2 = 0.88$). The viscosity of the solution estimated by extrapolation at the onset of spherulite nucleation ($\beta_{Mg} = 1.4$), is about 111 Pa$\cdot$s, comparable to melt-grown spherulite viscosities \cite{magill2001review}. Such a viscosity increase suggests a sol-gel transition, with diffusion becoming the rate-limiting step for crystal growth in this medium. Our SEM images indeed reveal a gel-like residue surrounding sodium sulfate spherulites (Fig. \ref{fig:SEM}). The combination of divalent ions (Mg$^{2+}$ or Fe$^{2+}$) and Na$^+$ appears crucial in achieving this exponential increase in viscosity. It enables diffusion-limited growth without completely impeding crystallization of the monovalent ion.
Our results show that viscosity variations and viscosity gradients drive morphological transitions: (i) a higher viscosity results in the growth of small, smooth spherulites (similar to $x_{Mg}$=0.25 in Fig. \ref{fig:steiger}a); (ii) a lower viscosity results in the growth of larger spherulites and in nanocrystals growing into more faceted (bladed) microcrystals.

\subsection*{Evaporation rate effects shape evolution}

\begin{figure*}
    \centering
    \includegraphics[width=\textwidth]{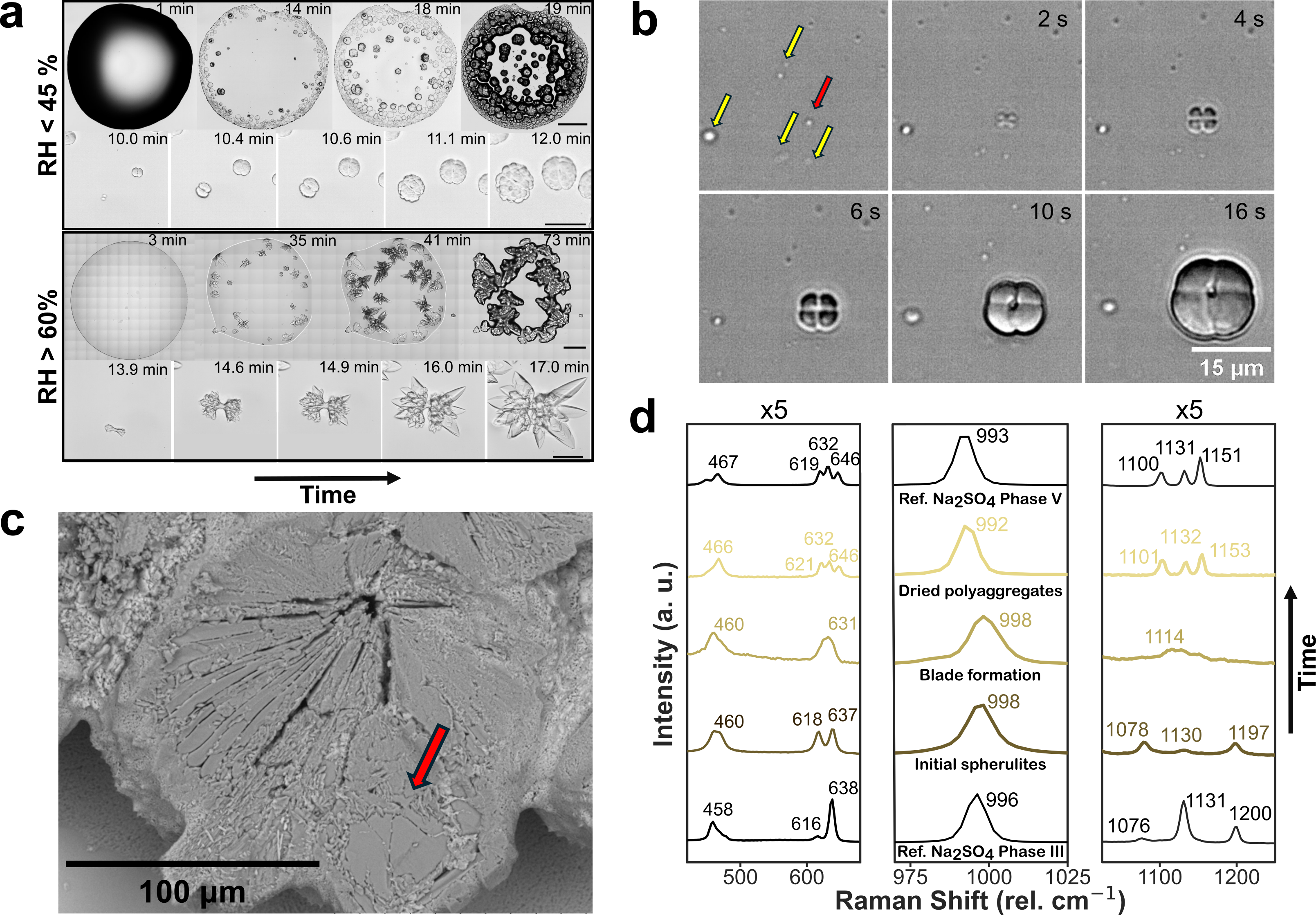}
    \caption{ \textbf{Impact of evaporation rate on the nucleation and growth of sodium sulfate spherulites}.\textbf{a} Typical crystallization stages of a 1 $\mu$L droplet with molar fraction $x_{Mg} = 0.12$  at fast (RH $\leq$ 45\%, top) and slow (RH $\geq$ 60\%, bottom) evaporation rates. Scale bars are 50 $\mu$m.
    \textbf{b} Optical microscope images of a spherulite growing from  1 a solution with molar fraction $x_{Mg} = 0.12$. Arrows highlight $\mu$m-sized pre-nucleation clusters before the onset of spherulitic growth. The red arrow points to how one of these preclusters grows into a spherulite of 15 $\mu$m over time. \textbf{c} Scanning electron microscopy image of a spherulitic polyaggregate after complete dehydration at RH=63\%. The red arrow highlights the growth history of a blade that developed from the spherulite.  \textbf{d} Raman spectra of late-stage evolution of spherulites in a $x_{Mg}$ = 0.12 droplet under slow evaporation conditions (RH $\geq$ 60\%). These measurements reveal that the initial spherulites correspond to phase III, bladed crystals formed on the spherulites originate from an unidentified phase, and the final polyaggregates recrystallize into phase V after complete dehydration. The spectra are accompanied by reference spectra of Na$_{2}$SO$_{4}$ phase III and phase V (shown in black).}
    \label{fig:RH}
\end{figure*}

To understand the influence of evaporation rate on spherulitic growth of sodium sulfate we let droplets evaporate at relative humidities (RH) between 30\% and 80\%. Our results (Fig. \ref{fig:RH}) demonstrate that spherulitic growth of sodium sulfate occurs across all tested RHs in both Fe- and Mg-containing solutions ($x = 0.12$). However,  the number of nuclei and the final morphology exhibit a strong dependence on evaporation rate (see also Fig. S\ref{fig:Supp_Drop_FeMg_RH}).
Fig. \ref{fig:RH}a illustrates the drying process of 1 $\mu$L droplets with $x_{Mg}$ = 0.12 at low and high RH. In both cases, the first spherulites generally nucleate near the contact line (where the droplet evaporation is strongest) and propagate inward, toward the droplet center.
Microscale phase-contrast microscopy analysis reveals that once high supersaturation is achieved by evaporation,  first tiny spherical droplet-like clusters ($\sim$ 1 $\mu$m in size) appear near the contact line, which are stable for several minutes \cite{bonn2016multistep}. These  spherulites subsequently grow following a two-step nucleation mechanism in the viscous, supersaturated solution (Fig. \ref{fig:RH}b). This non classical nucleation mechanism also explains the large number of spherulites precipitating in the evaporating droplets.
At high evaporation rates (RH $\leq$ 45\%, top panel), larger number of nuclei are generated compared to lower rates, resulting in the growth of smaller spherulites. At lower evaporation rates (RH $\geq$ 60\%, bottom panel), nanocrystals within the initially precipitated spherulites have sufficient time to grow to their equilibrium configuration, forming clusters of highly faceted crystals with different morphologies. As spherulitic growth progresses, it depletes the surrounding ion concentration, reducing the supersaturation in the surrounding medium. Thus, smooth faceted crystalline phases can form on existing spherulites (Fig. \ref{fig:RH}c), unless further growth is inhibited by an increase in solution viscosity (resulting in the formation of an amorphous state) or terminated by complete evaporation, thereby preserving the metastable spherulitic structure.

The Raman spectra obtained at each stage of growth provide key insights into the morphological transformation of spherulites (Fig. \ref{fig:RH}d). Comparison with reference spectra (shown in black) reveals that the initial spherulites consist of phase III  (needles), confirming previous results \cite{bonn2016multistep}. The Raman spectra of bladed crystals do not correspond to any known reference spectra in the literature and cannot be identified. These crystals are probably a metastable state, slowly transforming to a more stable state by cracking and recrystallizing (Fig. S\ref{fig:SEM_Blades}); the Raman spectra after recrystallization correspond to the stable phase V (thenardite), confirming the hypothesis that spherulites evolve toward polycrystalline aggregates with thermodynamically stable crystalline phases.
These findings demonstrate that spherulitic precipitation in highly viscous mixed sulfate solutions at very high supersaturation is a metastable intermediate state under non-equilibrium conditions. When the evaporation rate is slow and the growing spherulites have access to ions in the solution around them, they can evolve toward the more stable polycrystalline phase that is characteristic of the given salt---in our case, thenardite. The diverse spherulitic morphologies, ranging from smooth spherical to flower-like and/or needle-like structures, serve as a record of the system's growth history.

\subsection*{From fundamental understanding to morphological control of spherulitic growth}

The origin of spherulitic growth has been a subject of debate between classical and two-step nucleation mechanisms \cite{beck_spherulitic_2010, sand_crystallization_2012}. Wang \textit{et al.} observed that calcium sulfate first agglomerates in amorphous clusters before crystallizing into nanocrystals \cite{wang2013confinement}. Similar observations in other systems support the two-step nucleation mechanism: Vallina \textit{et al.} demonstrated that rare-earth carbonate spherulites initiate through amorphous precursor nucleation while Freitas \textit{et al.} proposed a two-step nucleation mechanism involving amorphous phase confinement \cite{vallina_role_2015, freitas2021crystallization}. Our results confirm the two-step nucleation process for the growth of sodium sulfate spherulites at high supersaturation from various mixed sulfate salt mixtures; sodium sulfate precipitates as spherulites from dense liquid droplets in a viscous highly concentrated medium instead of a classical needle-like dentritic growth from a salt solution. 

Models simulating spherulitic growth and the vast majority of experimental case studies suggest that non-crystallographic branching requires impurities in the system \cite{granasy2005growth,Shtukenberg2012spherulites}. Some research contradicts this, reporting successful spherulitic growth from pure compounds \cite{magill2001review, ryschenkow1988bulk, bisault1991spherulitic}. 
Here, we show that divalent (metal) ions in the solution specifically have the ability to induce spherulitic growth of the monovalent salt, by introducing complex competition in ion arrangement as they induce viscosity changes in the highly concentrated medium during evaporation, disrupting classical growth pathways and enabling non-crystallographic branching. This process results in self-organized clusters composed of small crystals with slight mismatches in orientation. 

\begin{figure}[h!]
    \centering
    \includegraphics[width=\linewidth]{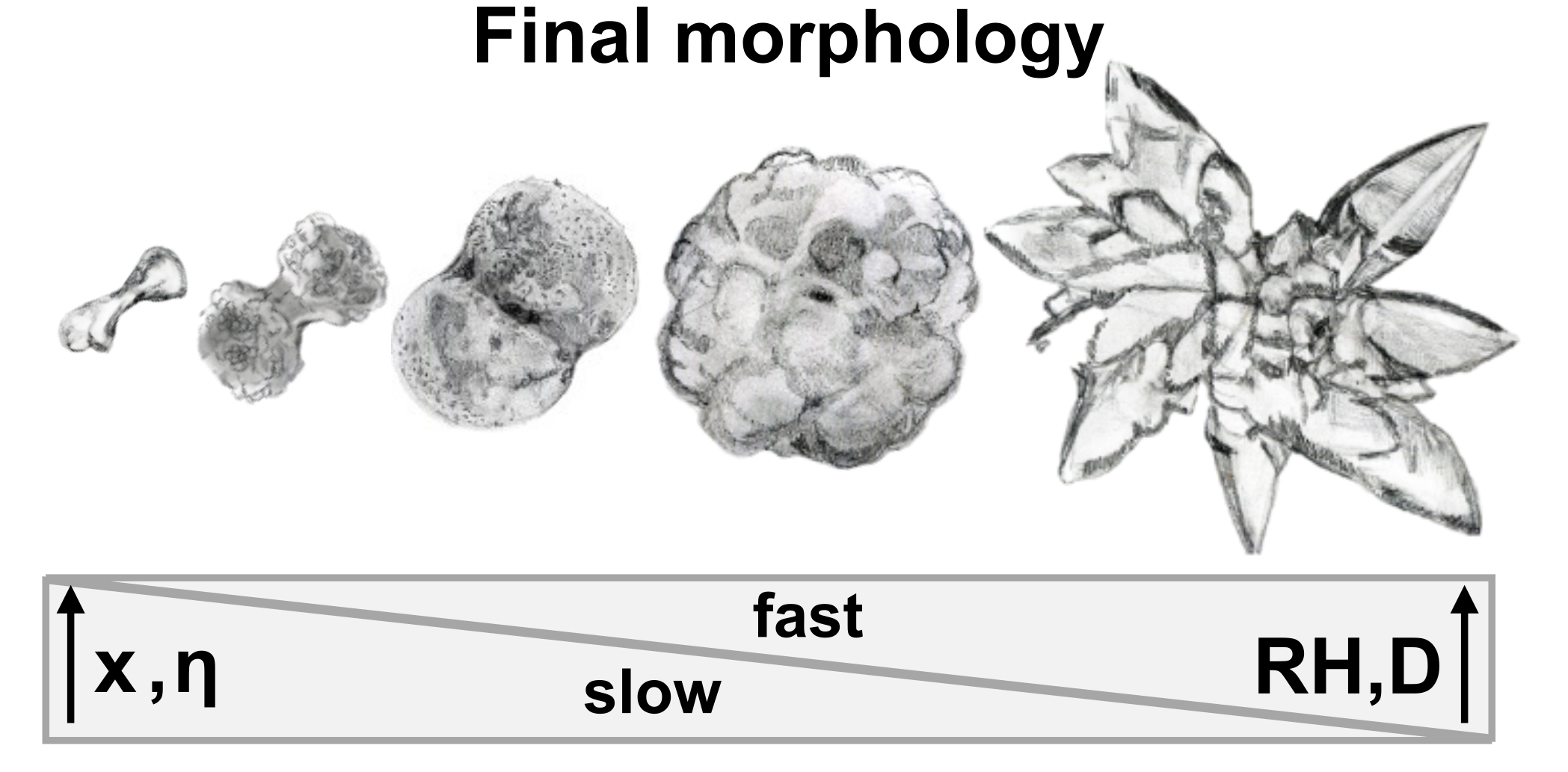}
    \caption{\textbf{Schematic representation of morphological transitions during spherulitic growth (not to scale)}. The spherulitic morphology evolves from compact forms to highly faceted structures as a function of experimental parameters such as relative humidity (RH), diffusion coefficient $D$, viscosity $\eta$, and the molar fraction of divalent ions $x$. The transition reflects the interplay between growth kinetics and environmental conditions, with slower growth resulting in rough and large morphologies and faster growth resulting in smooth spherical  structures with higher density.}
    \label{fig:finalmorphology}
\end{figure}

Figure \ref{fig:finalmorphology} illustrates the morphological transitions during spherulitic growth, highlighting the key experimental factors that influence the final morphology. Higher molar ratios $x$ of divalent ions lead to more viscous solutions, which in turn results in densely grown, smooth spherulites, and which prevent or delay the growth of other stable phases. At low $x$, depending on the RH, the growth of the more stable phase can be initiated at various stages of branching, including sprouting needles, premature (open) spherulites, and fully grown spherulites, leading to a wide range of final morphologies.  

These insights allow us to successfully grow sodium sulfate spherulites, not only from evaporative microdroplets but also from high-$x$ bulk solutions in Petri dishes, resulting in millimeter-sized spherulites (Fig. S\ref{fig:macrospherulites}). These findings shed light on the conditions under which natural peculiar-shaped crystals form, such as the gypsum sulfate spherulite in Fig. \ref{fig:CaSO4crystal}, or desert roses that form from mineral-rich water in the presence of silica (sands) in arid regions where the evaporation rate is high. It also offers practical strategies for tuning spherulitic morphologies in experimental settings. This capability has significant potential for innovative applications where specific crystal architectures are desired.

\section*{Conclusion} 

We report on the formation of Na$_{2}$SO$_{4}$ spherulites in evaporating salt mixtures. While pure sodium sulfate solutions typically yield regular polyhedral crystals upon evaporation, we found that polycationic sulfate salt mixtures with divalent ions induce spherulitic growth of sodium sulfate as polycrystalline aggregates of the metastable phase III by a two-step nucleation process.  Our analysis reveals specific ranges of molal fractions conducive to spherulitic growth: 0.03 < $x_{Fe}$ < 0.29 and 0.04 < $x_{Mg}$ < 0.38. To our knowledge, the ability of sodium sulfate to grow as spherulites has not been reported in the literature.
We also charted three stages in the morphological evolution of spherulites: (1) formation of micron-sized pre-nucleation clusters, (2) nucleation followed by growth from premature to fully developed spherulites, and (3) transformation into more stable polyhedral forms. This sequence underscores the metastable nature of spherulites, formed under non-equilibrium conditions, and their capacity to evolve as the supersaturation decreases during growth. This remarkable transformation highlights the profound impact of solution composition on crystallization pathways and the interplay between thermodynamics, kinetics, and morphology. 
We calculated the solubility diagrams of the ternary systems, Na$_{2}$SO$_{4}$--FeSO$_{4}$--H$_{2}$O and Na$_{2}$SO$_{4}$--MgSO$_{4}$--H$_{2}$O, using a Pitzer ion interaction model. Based on these diagrams, we successfully determined that the sulfate concentrations at the onset of spherulitic growth significantly exceed the solubility limit of phase III, with supersaturation levels reaching on average $\beta _{Mg}=1.4 \pm0.3$ and $\beta _{Fe}$>3.3. While the equilibrium phase diagrams are important tools to predict the supersaturation and the nature of the precipitating crystals, they cannot take into account the kinetics parameters and the properties of the solutions inducing the precipitation of the metastable phase III, as polycrystalline clusters, i.e. spherulites. Notably, we observed an exponential increase in viscosity with saturation in Na$_{2}$SO$_{4}$--MgSO$_{4}$ systems, reaching approximately 111 Pa$\cdot$s at the onset of spherulite precipitation. 

The formation of an amorphous phase at high molar fractions of divalent ions in the binary salt mixture leads to increased viscosity, delaying nucleation and inhibiting rapid spherulite growth. This phenomenon establishes diffusion as the limiting factor for growth, offering a novel perspective on the role of solution rheology in crystal morphogenesis. Our kinetic analysis reveals that spherulitic sodium sulfate crystals in both sulfate mixtures follow a diffusive growth mechanism, characterized by L $\sim$ $\Delta$ $\sqrt(t)$, where $3.9 \le 
\Delta \le 9.6$. The growth dynamics are governed by the interplay between diffusion rate $D$, viscosity $\eta$, and integration rate (G), dependent on supersaturation. 
We demonstrated that precise control over the final spherulitic morphology is achievable by modulating the molar fraction of divalent ions, evaporation rate, and geometric constraints.

Our results reveal the mechanisms of one of nature's most elegant self-assembly processes from salt mixtures. Understanding spherulitic growth mechanisms could revolutionize fields ranging from materials engineering and pharmaceutical formulation to geological modeling and biomineralization. By unraveling the principles of crystal self-organization, we pave the way for the rational design of complex materials with tailored structural and functional properties.

\section*{Materials and methods} 

\subsection*{Modeling of thermodynamic equilibria}
The thermodynamic equilibria in this study are calculated based on the following equations:
For a salt M${\nu_M}$X${\nu_X}$ $\cdot$ $\nu_0$H$_{2}$O, the equilibrium constant K$_{MX}$ of the dissolution reaction is: \begin{equation}
\begin{aligned}
\ln K_{M X} =& 
\nu_M \ln m_M+\nu_X \ln m_X+\nu_M \ln \gamma_M \notag \
+ \\ &\nu_X \ln \gamma_X+\nu_0 \ln a_{\mathrm{w}},
\end{aligned}
\end{equation} 
where m$_M$, m$x$, $\gamma_M$, and $\gamma_X$ are molalities and activity coefficients of cations and anions, respectively. Water activity a$_{\mathrm{w}}$ is defined as: \begin{equation}
\ln a_{\mathrm{w}} = -\phi M_{\mathrm{w}} \sum_i m_i,
\end{equation} with $\phi$ the osmotic coefficient and M$_{\mathrm{W}}$ the molar mass of water (M$_{\mathrm{W}}$ = 1.801528 $\times$ 10$^{-2}$ kg mol$^{-1}$). Further details about this model can be found in Steiger \textit{\textit{et al.}} \cite{steiger2008improved}.
The Pitzer model parameters for Na$_2$SO$_4$(aq), MgSO$_4$(aq), and Na$_2$SO$_4$--MgSO$_4$--H$_2$O were reported earlier~\cite{steiger2008improved,steiger2011decomposition}. The model parameters for FeSO$_4$(aq) are based on the model of Kobylin et al.~\cite{kobylin2011thermodynamic} with parameters listed in Talreja-Muthreja and Steiger \cite{talreja2025thermodynamics}. The ternary parameters for the system Na$_2$SO$_4$–-FeSO$_4$–-H$_2$O were determined from available solubility data. The parameters for the calculations at 21\textdegree C are $\theta_{\mathrm{Na,Fe}} = 0$ and $\psi_{\mathrm{Na,Fe,SO_4}} = 0.00234$. 

\subsection*{Salt mixture preparation} 
Na$_{2}$SO$_{4}$--FeSO$_{4}$--H$_{2}$O and Na$_{2}$SO$_{4}$--MgSO$_{4}$--H$_{2}$O mixtures were prepared by dissolving thenardite (Na$_{2}$SO$_{4}$) with either FeSO${4}\cdot 7$H$_{2}$O or MgSO${4}\cdot 7$H$_{2}$O in Millipore water (Sigma Aldrich, purity > 99 \%). The Na$_{2}$SO$_{4}$ concentration was set to 0.7 times the mirabilite solubility. Molar fractions $x$ of FeSO$_{4}$ or MgSO$_{4}$ were calculated using Eq. (\ref{eq:Y}). Salts were weighed with 0.05 g precision and dissolved in water by stirring for 30 minutes. 

\subsection*{Evaporation experiments} Droplets (1 $\pm$ 0.2 $\mu$L) were deposited on hydrophilic glass slides and observed under an inverted Leica DM-IRB microscope at $21 \pm 1 ^\circ$ C and 30-80\% RH. Crystallization was recorded at 0.5 fps, and spherulite growth rates were measured. Additional experiments in 0.5 $\times$ 0.5 mm$^2$ microcapillaries determined supersaturation at spherulite precipitation onset. 

\subsection*{Supersaturation measurements}
To quantify the supersaturation at the onset of sodium sulfate spherulites nucleation in both Fe and Mg solutions, initial volumes of mixed salt solution ($x = 0.12$) are confined in a microcapillary and allowed to evaporate in a controlled climatic chamber at $T$ = 21 $\pm$ 1$^\circ$C and RH = 50\% \cite{desarnaud2014metastability}. The change in volume as a function of evaporation time $t$ is subsequently followed by recording the displacement of the two menisci while simultaneously visualizing the onset of spherulite nucleation in the solution with an optical microscope coupled to a CCD camera. This measurement yields the molality at the onset of nucleation (see details in Supplementary C). 

\subsection*{Viscosity measurements}
Viscosities of solutions with varying water content ($x_{Mg}$ = 0.12, 0, and 1) were measured using an Anton Paar MCR301 rheometer. Samples were heated to 30$^\circ$C for homogeneity, then cooled to 21$^\circ$C for measurements. Viscosity was recorded at a shear rate of 900 s$^{-1}$ until reaching a plateau. The experimental protocol consisted of dissolving salts at elevated temperature ($T$ = 30$^\circ$C), followed by cooling to experimental conditions ($T$ = 21 $\pm$ 1$^\circ$C); subsequently we measured viscosity at different saturation stages, and define the evaporation stages were by the saturation of the Mg solution: $\beta_{Mg} =\frac{m}{m^{*}}$.

\subsection*{Structural and chemical characterization} Raman confocal microspectroscopy (WITec Alpha 300 R, 532 nm laser) provided chemical imaging during evaporation and crystallization. Scanning Electron Microscopy (FEI Verios 460) investigated spherulite structure at nano- to micro-scale, with samples coated in 80 nm gold particles and analyzed at 10 kV and 100 pA.

\section*{Acknowledgments}
We thank Julia Hooy for her help with the visual schematic. 

\section*{Authors contribution statement}
T.H., S.L., R.C., I.Y. and N.S. designed and performed experiments and analyzed the data. P.K. was involved in some of the experiments.  M.S. performed the modeling. T.H., D.B., M.S. and N.S. contributed to the final version of the manuscript.

\section*{Additional information}
The authors declare no competing interest.

\newpage
\bibliography{scibib}


\clearpage

\section*{Supplementary}
\input{Supplementary}

\end{document}

%% file: Supplementary.tex
\subsection{Solubilities of pure components}

\begin{figure}[!ht]
    \centering
    \includegraphics[width=\linewidth]{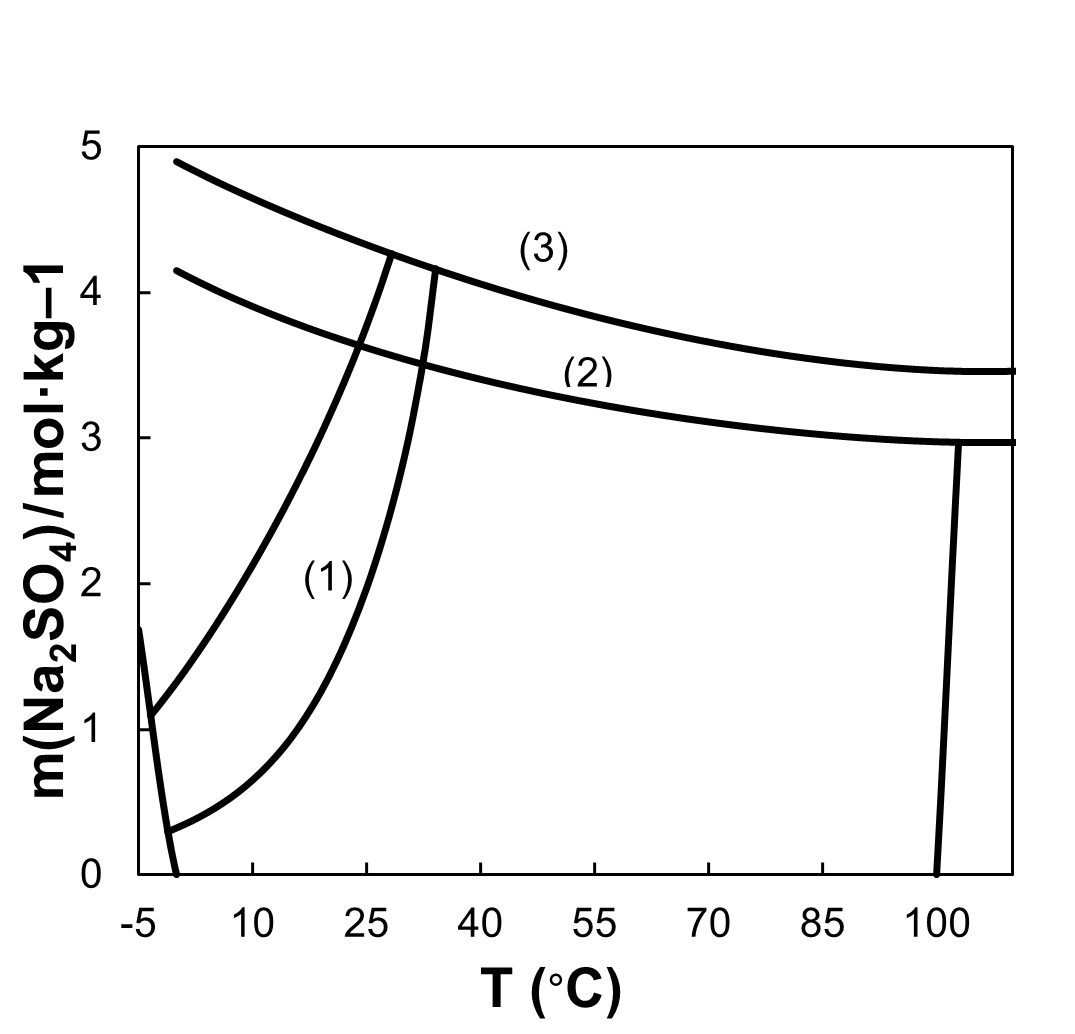}
    \caption{\textbf{Thermodynamic equilibria of sodium sulfate.} Solubility diagram of Na$_{2}$SO$_{4}$--H$_{2}$O  \cite{steiger2008crystallization}, including (1) Na$_{2}$SO$_{4}\cdot$10H$_{2}$O, (mirabilite), (2) Na$_{2}$SO$_{4}$ phase V (thenardite), and (3) Na$_{2}$SO$_{4}$ phase III.}
\label{fig:Appendix5_phasediagrams_literature}
\end{figure}

The solubility curves of sodium sulfate in water are previously reported by Steiger \textit{et al.} \cite{steiger2008crystallization}, and are presented in Fig.~\ref{fig:Appendix5_phasediagrams_literature}. 
The phase diagram includes the common mineral forms of sodium sulfate; (1) mirabilite, specifically the decahydrate form (hydrated by 10 water molecules) and (2) the anhydrous rhombohedral stable phase V called thenardite. The metastable dendritic/needle-like phase III (3) is included as well.
We summarize the solubilities of the pure salts in this study at $T$=21$^{\circ}$C in Table \ref{Table_ThermoSalt}, that are all calculated with the same model as the thermodynamic equilibria in Fig. \ref{fig:steiger} and Fig. \ref{fig:Appendix5_phasediagrams_literature}.

\begin{table}[!ht]
\caption{\textbf{Water solubilities of sulfate compounds at $T$ = 21 $^{\circ}$C}.}
\centering
\begin{tabular}{|l|l|}
\hline
Compound & \begin{tabular}[c]{@{}l@{}}Water solubility\\ (mol $\cdot$ kg$^{-1}$)\end{tabular} \\ \hline
\begin{tabular}[c]{@{}l@{}}Mirabilite\\ (Na$_{2}$SO$_{4}\cdot$ 10H$_{2}$O) \end{tabular} & 1.46 \\ \hline
\begin{tabular}[c]{@{}l@{}}Thenardite, phase V\\ (Na$_{2}$SO$_{4}$)\end{tabular} & 3.69 \\ \hline
\begin{tabular}[c]{@{}l@{}}Phase III \end{tabular} & 4.41 \\ \hline
\begin{tabular}[c]{@{}l@{}} FeSO$_{4} \cdot$ 7H$_{2}$O\end{tabular} & 1.79\\ \hline
\begin{tabular}[c]{@{}l@{}} MgSO$_{4} \cdot$ 7H$_{2}$O\end{tabular} & 2.88\\ \hline
\end{tabular}
\label{Table_ThermoSalt}
\end{table}

\subsection{Raman spectra of sodium sulfate spherulites and reference phases}
\begin{figure*}
    \centering    
    \includegraphics[width=0.75\textwidth]{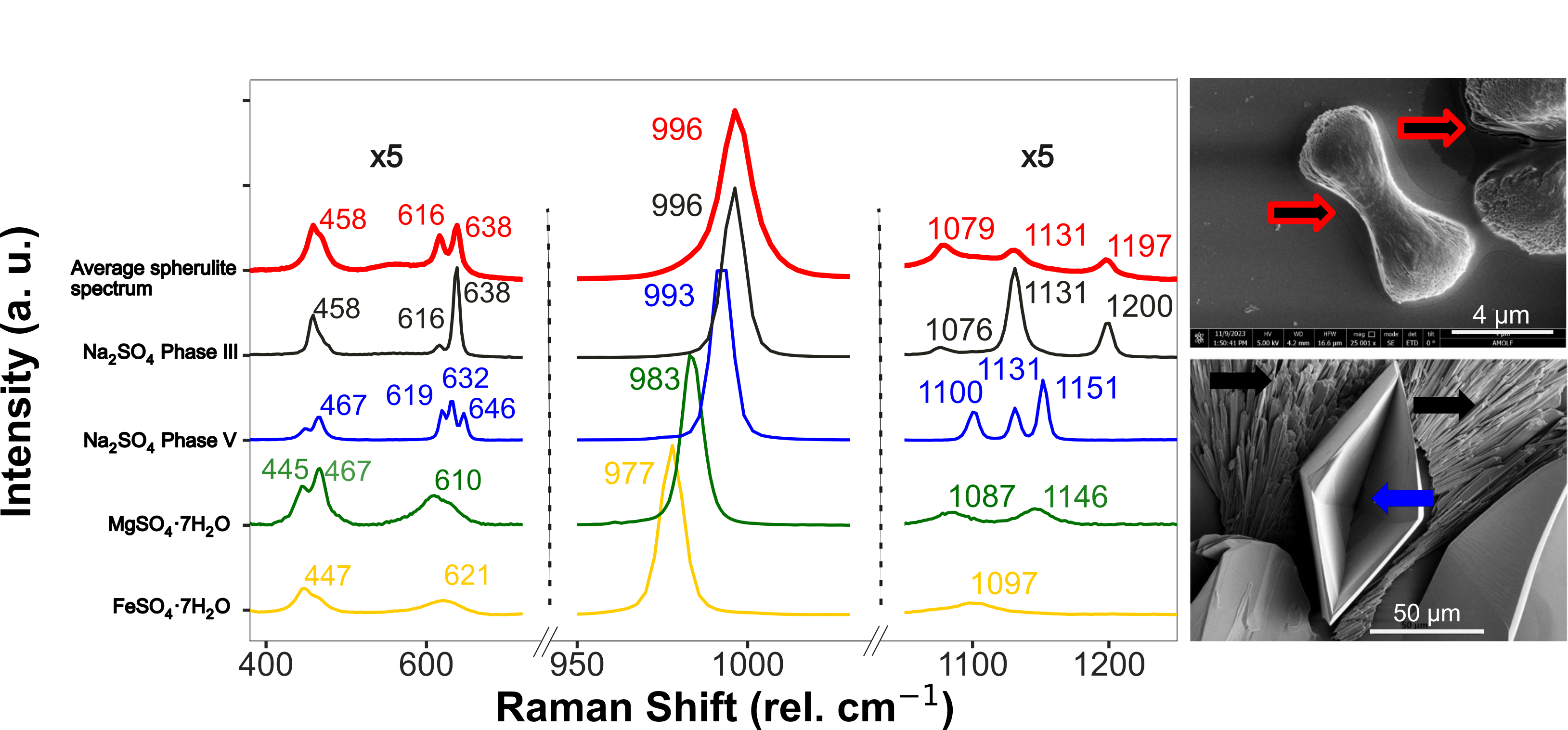}
    \caption{
    \textbf{Average Raman spectra of spherulites compared with (Na, Mg, Fe) sulfate phases.}
   Left: (red) average Raman spectra of 7 spherulites, (black) Raman spectra of Na$_{2}$SO$_{4}$ phase III, (blue) Na$_{2}$SO$_{4}$ phase V, (green), MgSO$_{4}\cdot$7H$_{2}$O, (yellow), FeSO$_{4}\cdot$7H$_{2}$O.  Left and right regions are 5$\times$ magnified.  Right: SEM image of (premature) Na$_{2}$SO$_{4}$ spherulite crystallized from $x_{Mg}$ = 0.12 sulfate mixture (top), compared with stable rhombohedral thenardite and needle-like metastable phase III from pure Na$_{2}$SO$_{4}$ solution (bottom, reproduced from \cite{bonn2016multistep}).}
    \label{fig:raman_spherulites}
\end{figure*}

 Fig.~\ref{fig:raman_spherulites} shows anhydrous phase morphologies after drying pure Na$_{2}$SO$_{4}$ solution, with metastable phase III (needle-shaped) transforming into stable phase V (rhombohedral). At room temperature (21 $\pm$ 1$^\circ$C), mirabilite is the stable hydrated form in equilibrium with saturated solution. Hydrated crystals dehydrate spontaneously to phase V in unsaturated air (RH$^{dehydration}_{eq}$ = 65\%) at room temperature. Fig.~\ref{fig:raman_spherulites} compares the anhydrous phases after drying pure Na$_{2}$SO$_{4}$ solution (metastable needle-shaped phase III and stable rhombohedral shape phase V) with the spherulites by Raman confocal microspectroscopy and SEM images. Raman confocal microspectroscopy reveals that the spherulites consist of anhydrous sodium sulfate nanocrystals in the needle-shaped phase III (thenardite). The overlapping Raman peaks (458, 616, 638, 996, 1076-1079, 1131, 1197-1200 cm$^{-1}$) between the average spectra of 5 spherulites and the reference spectrum phase III confirm this composition.

 \subsection{Exploratory experiments with varying divalent ions} \label{sec:Supp_divalentions}

\begin{figure}[!ht]
    \centering    
    \includegraphics[width=\linewidth]{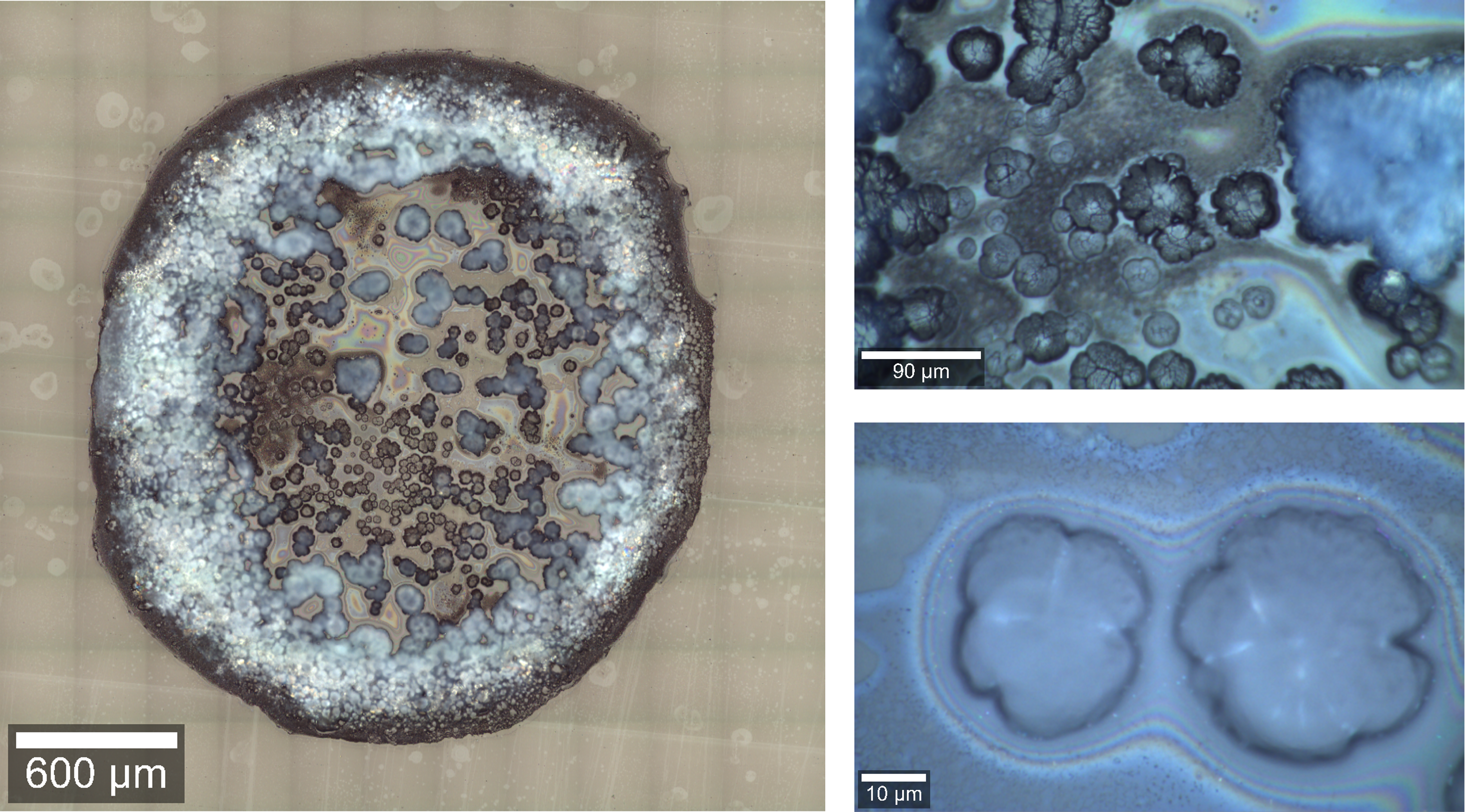}
    \caption{\textbf{Microscopic images of Na$_{2}$SO$_{4}$ spherulites in presence of $Cu^{2+}$ ions}. The spherulites crystallized from a 1 $\mu$L droplet Na$_{2}$SO$_{4}$--CuSO$_{4}$--H$_{2}$O  solution.}
    \label{fig:cu_spherulites}
\end{figure}

 Exploratory experiments involving the substitution of Fe$^{2+}$ and Mg$^{2+}$ with other divalent salts such as Cu$^{2+}$ and Zn$^{2+}$ also led to similar sodium sulfate spherulite precipitation, suggesting that the mechanism of spherulitic growth is not system-specific and can occur in different mixtures of divalent sulfate salts. An example of sodium sulfate spherulite precipitation in the presence of Cu$^{2+}$ is presented in Fig. \ref{fig:cu_spherulites}. 
 
\subsection{Determination of the supersaturation} \label{sec:Supp_supersaturation}

\begin{figure}[ht!]
    \centering
    \includegraphics[width=\linewidth]{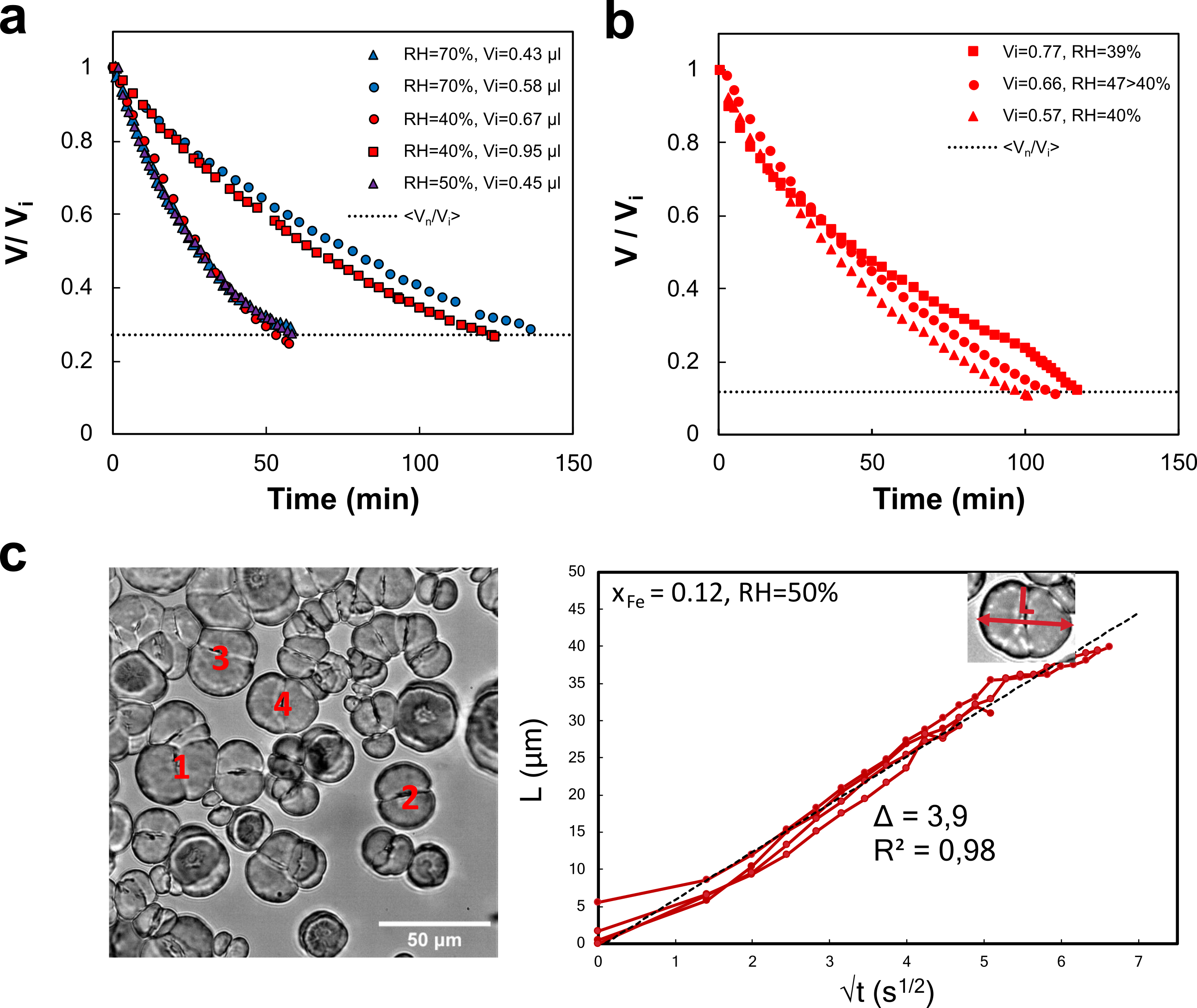}
    \caption{\textbf{a} Volume fraction of $x_{Mg}$=0.12 solution \(\frac{V}{V_i}\) in microcapillaries until nucleation. \textbf{b} Volume fraction of $x_{Fe}$=0.11 solution \(\frac{V}{V_i}\) in microcapillaries until nucleation. \textbf{c} Growth rate of sodium sulfate spherulites in Fe solution (RH=50\% and $x_{Fe}=0.12$). The length $L$ of five spherulites (labelled 1, 2, 3, 4, and 5 in the image) is plotted as a function of the square root of time.  The linear trend (slope $\Delta$) indicates that the growth of the spherulites is controlled by diffusion.  }
    \label{fig:Supp_Fe_growth}
\end{figure}

The supersaturation at the onset of spherulitic growth was determined by evaporating the $x_{Mg}$ and $x_{Fe}$ solutions in microcapillaries of width 0.5 $\times$ 0.5 mm$^{2}$, similarly to Desarnaud \textit{et al.} \cite{desarnaud2018hopper, desarnaud2014metastability}. Experiments were performed under controlled evaporation ($T$=21 $\pm$1\textdegree C and RH=50\%) of small volumes (0.5 to 1 $\mu$L) of initially undersaturated $x_{Mg}$ and $x_{Fe}$ solutions.
The temporal evolution of the volume of the salt solutions was calculated from the dimensions of the capillary and the distance between the two menisci up to crystallization (Fig. \ref{fig:Supp_Fe_growth}(a,b)). 
The concentrations at the moment of precipitation were determined by the difference between the initial volume ($V_{0}$) and the volume at the moment the system starts to nucleate ($V_{n}$). This yields molar concentrations (number of moles solute per volume of solution) at the onset of nucleation: $c_{n}=V_{0}/V_{n}c_{0}$.

The conversion of the volume-based concentrations $c$ to molalities $m$ (number of moles per mass of water) and vice versa requires solutions densities \cite{shen2020crystallization}. In the Na$_{2}$SO$_{4}$--MgSO$_{4}$ mixtures, the densities were calculated using a volumetric Pitzer model \cite{linnow2009modeling} yielding an average molality of sulfate in the supersaturated solutions of 6.21$\pm$0.45 mol$\cdot$kg$^{-1}$. The sulfate molality in the starting solution was 1.43 mol$\cdot$kg$^{-1}$. The saturation concentration at equilibrium with respect to sodium sulfate phase III ($m_{0.12}^*$) can be defined from Fig. \ref{fig:steiger}  (4.6 mol$\cdot$kg$^{-1}$), yielding to  $\beta$= 1.4.
The same volumetric Pitzer model could not be used for the calculation of densities in the Na$_{2}$SO$_{4}$--FeSO$_{4}$ for two reasons. First, the model does not include Fe(II) salts. Second, the concentrations of the supersaturated solutions were so high that no reliable densities for the highly supersaturated Na$_{2}$SO$_{4}$ solutions can be calculated. Available density data for Na$_{2}$SO$_{4}$ extend to $m$= 15 mol$\cdot$kg$^{-1}$ \cite{tang1994water}. An equation for FeSO$_{4}$ is available to $m$= 10 mol$\cdot$kg$^{-1}$ \cite{konigsberger2009measurement} and extrapolates reasonably. Thus, the upper limit for reliable density calculations is a total sulfate concentration of $m_{S}$= 15 mol$\cdot$kg$^{-1}$. Densities of mixed solutions can be calculated using a linear mixing rule:
\begin{equation}
d= (1-x_{Fe})d_{1}+x_{Fe}d2
\end{equation}
where $d_{1}$ and $d_{2}$  are the densities of the respective binary solutions at the sulfate molality of the mixed solution. $m_{S}$= 15 mol$\cdot$kg$^{-1}$ and $x_{Fe}$= 0.105 the resulting molar concentration is 9 mol$\cdot$L$^{-1}$ which is still slightly lower than the molarities observed in the microcapillary experiments (10–11 mol$\cdot$L$^{-1}$). Therefore, only a minimum supersaturation of $\beta$> 3.3 can be calculated ($m_{S}$/m$_{S}^{*}$).


\subsection{Growth measurements in Fe solution}
Similarly to what was done in the main text for Mg-containing solutions, the growth rate of spherulites in Fe-containing solution was determined at RH=50\% for four different spherulites, as shown in Fig. \ref{fig:Supp_Fe_growth}. The spherulite length $L$ (in mm) as a function of time is given by: $L = 3.9 \sqrt{t}$, which is similar to the case of spherulites growing near the droplet edges in Mg-containing solutions.

\subsection{Influence of the evaporation rate on the final morphology} \label{sec:Supp_RH}
\begin{figure}[!ht]
    \centering
    \includegraphics[width=\linewidth]{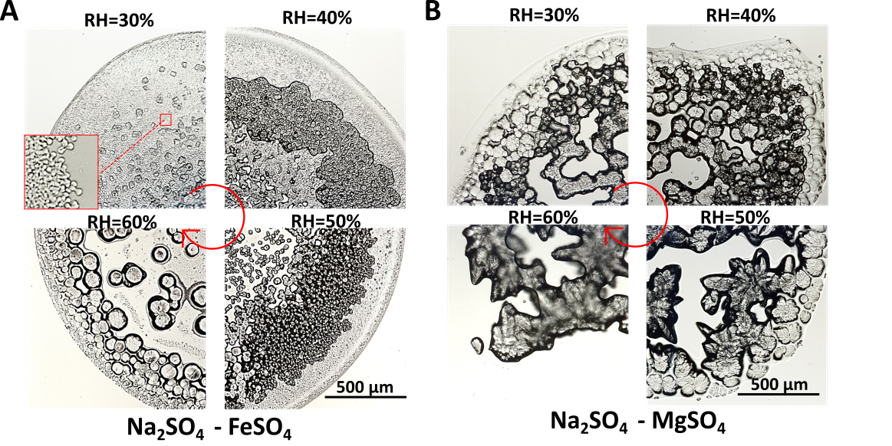}
    \caption{\textbf{Effect of evaporation rate on the morphological evolution.}  Experimental results of varying the RH for a Fe solution ($x_{Fe} = 0.12$) (\textbf{a}) and a Mg solution ($x_{Mg}=0.12$) (\textbf{b}). The RH increases in the clockwise direction (red arrow).}. 
    \label{fig:Supp_Drop_FeMg_RH}
\end{figure}

The growth of spherulites from 1 $\mu$L droplets of $x_{Fe} = 0.12$ solution under various relative humidities is compared to the case of $x_{Mg} = 0.12$ solution in Fig.~\ref{fig:Supp_Drop_FeMg_RH}.

\begin{figure}[ht]
    \centering
    \includegraphics[width=\linewidth]{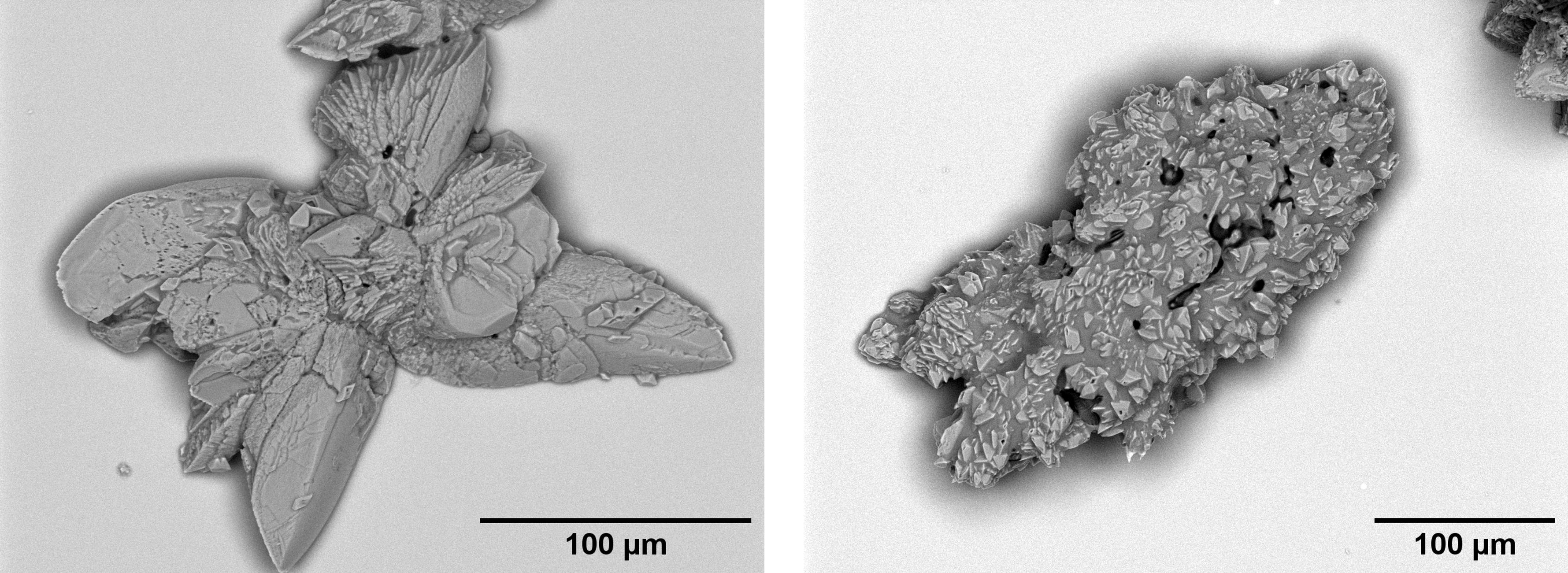}
    \caption{\textbf{Formation of blade-like structure in Mg solution.} SEM image of the late-stage formation of  ``blade” crystals on Mg-containing spherulites, before (left) and after (right) disintegration and recrystallization of thenardite ($x_{Mg} = 0.12$, RH = 60\%).}
    \label{fig:SEM_Blades}
\end{figure}

Similarly to the case of Mg-containing solutions, for high evaporation rates (low RH), the number of nuclei is much higher than in the case of low evaporation rates/high RH, with smaller spherulites forming toward the end of the drying process. Moreover, as for Mg, the spherulites that grow at the edge of the droplet, remain small and are stuck at an earlier stage of development than those precipitating in the center of the drop, independently of the relative humidity. Nonetheless, spherulites in droplets of the Fe- containing solution retain their spherical shape throughout their evolution whereas, when the evaporation is slow enough in the Mg-containing solution, faceted crystals shaped like blades grow from the spherulites. The blades formed on the spherulites in Mg-containing solutions are instable: over time, they crack and recrystallize as thenardite, as visible in the SEM pictures in Fig. \ref{fig:SEM_Blades}.
\vspace{0.2cm}

\subsection{Towards growth of macrospherulites} 

The rheological experiments of Fig. \ref{fig:growth_viscosity}b revealed that increasing solution volume 500-fold significantly impacts crystal morphology. To explore spherulite growth at larger scales, 10 mL of $x_{Mg}$ = 0.5 solution was introduced into hydrophobic and silicon environments with low relative humidity (RH < 10\%). The key observations
were that millimeter-sized spherulites successfully formed in both setups; SEM images confirm the characteristic spherulite structure. In addition, we observed that spherulites remained covered by a solidified gel layer and upside-down views revealed consistent small crystal aggregates. This experiment represents the first controlled laboratory demonstration of spherulite growth across multiple size scales, from microscopic to millimetric dimensions. The findings suggest that spherulite formation mechanisms are scalable and can be reproducibly controlled by carefully managing solution composition and environmental conditions. The ability to grow macro-spherulites opens new possibilities for understanding crystallization processes and potentially developing applications that leverage these unique polycrystalline structures.

\begin{figure}[!ht]
    \centering
    \includegraphics[width=\linewidth]{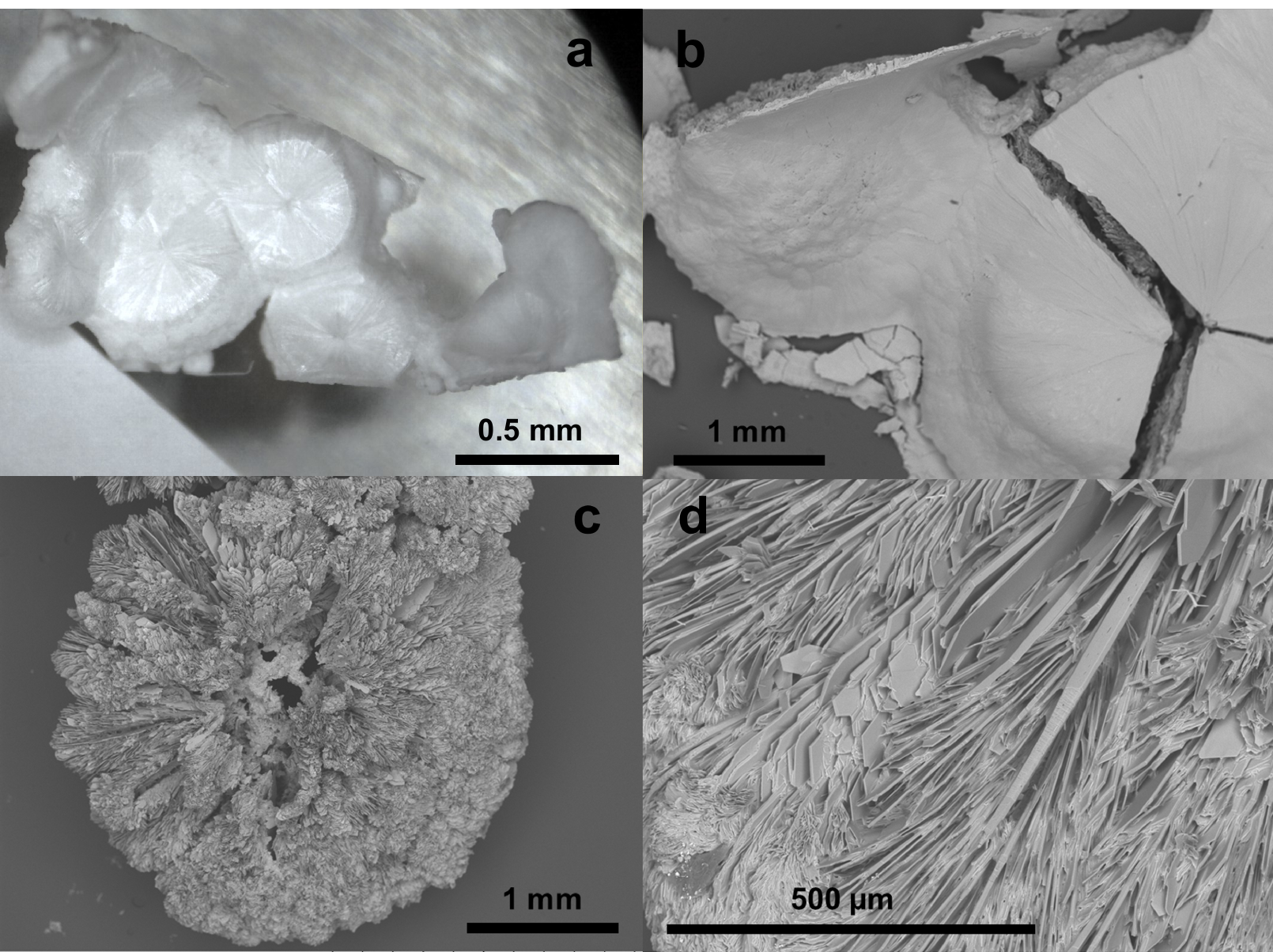}
    \caption{\textbf{Mm-sized spherulites crystallized from larger volumes.} \textbf{a} Bottom view from spherulites removed from a silicon holder. \textbf{b} SEM image of the same sample (top view). \textbf{c} SEM image of a single spherulite from the hydrophobic petridish (bottom view). \textbf{d} Zoomed-in image of the edge of the spherulite of panel c. }
    \label{fig:macrospherulites}
\end{figure}